\documentclass[aip,prl,amsmath,amssymb,amstext,reprint,citeautoscript,punctuation]{revtex4-1}

\usepackage{graphicx,xcolor,charter}

\begin{document}

\sloppy

\title{An unconventional mechanism for simultaneous high hardness and ductility in Mo$_{2}$BC}

\author{Aria Mansouri Tehrani}
\affiliation{Department of Chemistry, University of Houston, Houston, TX, USA}
\author{Amber Lim}
\affiliation{Department of Chemistry, University of Houston, Houston, TX, USA}
\author{Jakoah Brgoch}
\email{jbrgoch@uh.edu}
\affiliation{Department of Chemistry, University of Houston, Houston, TX, USA}
\affiliation{Texas Center for Superconductivity, Houston, TX, USA}


\begin{abstract}
Materials either have a high hardness or excellent ductility, but rarely both at the same time. Mo$_2$BC is one of the only crystalline materials that simultaneously has a high Vickers hardness and is also relatively ductile. The origin of this unique balance between hardness and ductility is revealed herein using first-principles stress-strain calculations. The results show an anisotropic response including a remarkable intermediate tensile strain-stiffening behavior and a two-step sequential failure under shear strain. The mechanism governing the mechanical properties more closely resembles ductile soft materials like biological systems or some polymer networks rather than hard, refractory metals. This unforeseen response is established by analyzing changes in the electronic structure and the chemical bonding environments under mechanical perturbation. Most importantly, the optimized structure under extreme strain shows the formation of a pseudogap in the density of states and dimerization of the structure's boron-boron zigzag chain. This leads to an enhancement of these strong covalent bonds that help delay the ultimate failure until higher than the anticipated strain. These results provide a platform for developing a new generation of high hardness, ductile materials by identifying compounds that form electronically metastable structures under extreme strain.

\end{abstract}

\maketitle

The balance of hardness and ductility is an essential materials consideration for any application.\cite{Ritchie2011,Wei2014, Kim2015, Ye2017} Cutting tools and wear resistant coatings require high hardness while brittleness can adversely affect their working lifespan due to the formation and propagation of cracks.\cite{Nakayama1988, Soler2018} Yet, developing new materials with both high hardness and excellent ductility is extremely challenging because nearly all crystalline solids are typically either hard and brittle or ductile and soft.\cite{Wang2014, Kim2015, Ye2017} Researchers have devised some approaches to overcome this trade-off. For example, the formation of ordered oxygen complexes can simultaneously enhance the strength and ductility in a class of high entropy alloys through a strain-hardening mechanism.\cite{Liu2018} Another research route to acquire high hardness and ductility is through the development of damage-tolerant architected materials. The various periodic arrangement of these materials can yield many unconventional mechanical properties, such as negative Poisson's ratio as well as materials with fine-tuned high strength and toughness.\cite{Pham2019} Varying the grain structure through extensive processing is also an avenue towards high hardness systems with moderate ductility by enhancing hardening mechanisms that work on longer length scales such as dislocation pinning\cite{Yuan2015}. Further, an increase in ductility has also been reported by engineering the stacking fault energies to limit the possible deformation modes in transition metals monocarbides.\cite{DeLeon2015}

There are significantly fewer examples of influencing the hardness/ductility based solely on modifying the structure or chemical bonding in simple crystalline solids. DFT calculations have shown that alloying the TiN with Mo and W enhances the toughness significantly.\cite{Sangiovanni2010} The system Mo$_2$BC also provides an opportunity to investigate a pseudo-layered crystal structure with an extremely high hardness (29 GPa \textit{via} nanoindentation\cite{Emmerlich2009}) that is moderately ductile. Indeed, given the chemical composition, this phase has a much higher than expected ductility based on its Pugh's ratio ($G/B$ = $\approx$0.57) falling on the border of the ductile regime, the positive Cauchy pressure\cite{Emmerlich2009}, and by in situ analysis of crack behavior during tensile testing.\cite{Djaziri2016} As a result, Mo$_{2}$BC is considered for application as a hard coating due to its high hardness, high fracture toughness, and relative ductility.\cite{Emmerlich2009, Soler2018} However, identifying the fundamental origins of plasticity in this phase or other metal-rich materials with complex crystal structures  remains elusive due to challenges in controlling sample preparation and testing conditions. Some progress has been made by developing small-scale testing methods such as micro-compression to study plasticity in hard materials. \cite{Korte-Kerzel2017} Additionally, first-principles calculations have provided insight into the structure-mechanical property relationships in materials, such as ternary carbides and borides.\cite{Zang2012, Jhi2001b, Yang2009}
Nevertheless, research applying these experimental or computational techniques to develop entirely new materials systems is limited owing to the difficulties associated with the experimental measurements and computational cost of simulating complex mechanical processes like indentation.\cite{MansouriTehrani2019} More recently, a machine learning based screening method provides an approach to identify potential high hardness materials \textit{a priori}.\cite{MansouriTehrani2018} For example, predicting the bulk ($B$) and shear ($G$) moduli (of 118,287 compounds) as a proxy for hardness recommended the further examination of a transition metal carbide, ReWC$_{0.8}$, and Mo$_{0.9}$W$_{1.1}$BC.\cite{MansouriTehrani2018}  Even though $B$ and $G$ are correlated to hardness and are capable of highlighting potential hard materials, the elastic moduli only pertain to small strains at equilibrium. On the contrary, indentation causes large strains and severe deformations at non-equilibrium conditions, which cannot be adequately explained by elastic moduli.\cite{Krenn2002, Ogata2002}  Thus, to fully understand the mechanisms involved in deformation once a new material has been identified with high-level techniques like machine learning, a more comprehensive analysis is necessary. 

One solution is to employ \textit{ab-initio} calculations of stress-strain behavior as a probe of materials at large strains. These calculations can be used to determine the incipient of plasticity (yield strength) as an upper bound for the strength of a defect-free crystal.\cite{Roundy1999, Krenn2002}  In fact, this idea has been used to explain the deformation response of Al and Cu metals\cite{Ogata2002}, determine the elastic instability in transition metal carbides and nitrides (HfC, TiN, and TiC)\cite{Jhi2001b}, study differences in the atomistic deformation modes of diamond, $c$-BN, and BC$_2$N,\cite{Zhanga} and understand how carbon content affects the strength of BC$_x$N.\cite{Zhang2008} Moreover, these calculations identified an unexpected strain-stiffening mechanism in Fe$_3$C and Al$_3$BC$_3$, which resembles the mechanical behavior of biological materials, such as skin.\cite{Jiang2012} More recently, two tungsten nitrides, hp4-WN and hp6-WN$_2$ were also predicted using crystal structure searching algorithms and suggested to have a hardness of $>$40 GPa, even in the asymptotic regime based on their high calculated ultimate stress. The extremely high hardness of these predicted nitrides is attributed to indentation compression induced strengthening and the strengthening of tungsten-tungsten bonds.\cite{Lu2017} Finally, stress-strain calculations revealed that a sequential bond-breaking mode was responsible for outstanding mechanical properties of the hypothetical $d$-BC$_3$.\cite{Zhang2015}

In our work to understand the mechanical properties and structural chemistry of transition metal borocarbides previously selected using machine learning, we identify an incredible and previously unobserved intermediate strain-stiffening mechanism. Mo$_2$BC experiences an electronic structure-mediated two-step failure during deformation. This response is a different failure mechanism compared to all other typical brittle, hard materials such as ReB$_2$ and diamond. The results more closely resemble biological systems like collagen, soft materials such as cross-linked rubber and polymer networks,\cite{Zhang2010,Zhang2004a, Meyers2009} or annealed AISI 1040 steel.\cite{Meyers2009} The mechanism of the observed intermediate strain-stiffening under tension stems from the formation of electronic pseudo-gap in the density of state under strain whereas a delayed failure during shear deformation occurs based on an initial dimerization of the B-B zigzag chains followed by the strengthening of the dimers. The evolution of this compound's electronic structure and its chemical bonding environments during the deformation result in the surprising ductility while maintaining the high hardness expected for a transition metal borocarbide. The findings of this research not only explain the fundamental mechanisms governing the mechanical properties in Mo$_2$BC, but it also provides a new perspective on the design of future mechanical materials by contradicting the notion of avoiding materials with pseudo-layered crystal structures.

\begin{figure} 
\includegraphics[width=3in]{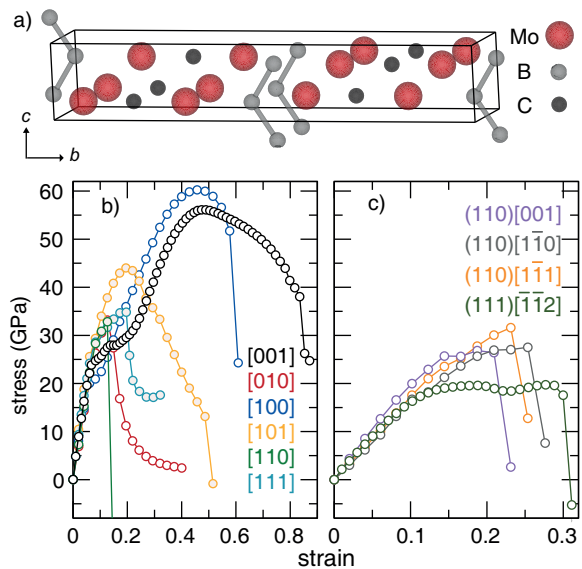} 
\caption{a) Crystal structure of Mo$_2$BC. b) Calculated stress-strain curves for various crystallographic directions during applied tension. c) Stress-strain curve for various shear systems.} 
\label{fig:1}
\end{figure}

The calculations were all performed using DFT within the framework of the Vienna \textit{ab-initio} simulation package (VASP), which employs a plane wave basis set with projector augmented wave (PAW) pseudopotentials.\cite{Kresse1996, Joubert1999,Blochl1994} The Perdew-Burke-Ernzerhof (PBE) generalized gradient functional was employed as an approximation for exchange and correlation.\cite{Perdew1996} An energy cutoff of 600 eV and a $k$-point mesh\cite{Monkhorst1976} of 10$\times$2$\times$10 for the irreducible wedge of the first Brillouin zone were used for all calculations. Electronic and structural optimizations were carried out with convergence criteria of 1$\times$10$^{-8}$ eV and 1$\times$10$^{-6}$ eV, respectively. The optimized crystal structure of Mo$_2$BC is illustrated in Fig.\,\ref{fig:1}a and contains infinite zig-zag chains of boron atoms along the [001] direction that alternate with layers of molybdenum atoms that center the basal plane of a square-based pyramid. The DFT optimized lattice parameters are $a$=3.096\,\AA, $b$=17.442\,\AA, and $c$=3.064\,\AA, which agrees with the experimentally refined values ($a$=3.086\,\AA, $b$=17.350\,\AA, and $c$=3.047\,\AA).\cite{Smith2002} For the ideal strength calculations, the unit cells are strained incrementally in 0.01\,\AA\ steps while the orthogonal lattice directions are fully relaxed.\cite{Zhang2015,Hao2015,Zhang2014} The compound is strained stepwise until reaching a plateau followed by a drop in the stress value.

To study the fundamental structural-mechanical properties of Mo$_2$BC, the stress-strain behavior was evaluated for numerous crystalline directions and planes. Typically, as the crystal is strained the stress is increased until reaching a maximum value, which is designated as the ideal (ultimate) tensile/shear stress (strength). The corresponding highest strain achieved prior to failure is called the ideal strain. Beyond this critical point, the crystal structure is either electronically or mechanically unstable and is held together only by repulsion forces. The computational results for Mo$_2$BC, plotted in Fig.\,\ref{fig:1}b, depict the uniaxial tensile stress-strain curves for the high symmetry directions. Similar to the anisotropy observed in this orthorhombic crystal structure, the mechanical response is also highly anisotropic. The lowest stress tolerated by the structure occurs for the [010] and [011] directions, which achieves a maximum of only $\approx$30\,GPa at 0.15 strain. These directions show the commonly observed linear increase in stress before reaching a plateau followed by a sudden drop at failure. The ultimate stress occurs for the [100] direction reaching $\approx$60\,GPa at 0.46 strain followed closely by [001] with a value of $\approx$56\,GPa at 0.49 strain.  These two directions are exceptionally separated from the other directions and show a comparatively much higher ultimate stress and strain.
The ultimate strength of these two directions in Mo$_2$BC is only slightly lower than superhard materials like ReB$_2$, which range from $\approx$\,48\,GPa for $(\bar{1}3\bar{1}0)$ to $\approx$76\,GPa for $(0001)$.\cite{Wang2009} 
A closer analysis of the [001] and [100] tension directions also highlights an entirely different mechanical behavior exhibiting an intermediate strain-induced stiffening that is more commonly observed in polymers, biological materials, and hydrogels. As these material are strained, they get stiffer.\cite{Erk2010, Jaspers2014}  The strain-induced stiffening in the crystalline material studied here is somewhat similar to cementite (Fe$_3$C) during shear deformation, which is the origin of its robust mechanical behavior; yet, cementite is significantly softer with Vickers hardness of only $\approx$10\,GPa compared with 29\,GPa \textit{via} nanoindentation for Mo$_2$BC.\cite{Jiang2012}  

Similarly, the mechanical behavior of Mo$_2$BC is also investigated under shear stress because indentation occurs through plastic deformation, which is governed by the movement of dislocations that only glide under shear strain.\cite{Gilman2003} For highly anisotropic materials such as MAX phases, the hardest slip plane is known to dictate indentation resistance whereas the softest slip system is responsible for the failure of the material. Identifying the softest or hardest slip planes require an exhaustive probe of the different slip systems. These calculations are provided in Table\,S1 of the supporting information. The highest ideal shear stress in Mo$_2$BC is achieved for the (110)[$\bar{1}$10] shear slip system with an ideal stress of 50 GPa while the highest ideal strain of 0.37 occurring for the (110)[1$\bar{1}$1] slip system. These properties are competitive with other high hardness materials such as ReB$_2$, where the highest ideal shear stress is 52\,GPa at 0.27 strain.\cite{Zang2012} Even though the ideal shear in ReB$_2$ is slightly higher than Mo$_2$BC, the ideal shear strain in Mo$_2$BC is significantly higher emphasizing its ductility. 

Additional calculations indicate the softest shear plane in Mo$_2$BC is (111)[$\bar{1}\bar{1}$2] with ideal stress of 20 GPa at 0.28 strain. For comparison, the softest slip system in OsB$_2$ during shear deformation is only $\approx$\,9 GPa\cite{Yang2008}, half of this borocarbide. However, the softest shear in ReB$_2$ is 36 GPa at 0.18 strain, indicating the hardness of Mo$_2$BC falls between these two diborides, in agreement with experimental observations. This shear stress-strain curve also shows a striking two-step progression before eventual failure. As illustrated in Fig.\,\ref{fig:1}c, the stress increases as a function of strain before beginning to decrease at approximately 0.19 strain. At this point, the material would be expected to fail; however, an atypical increase in the stress then occurs extending the strain until eventual ultimate failure at 0.3 strain with a stress of 20 GPa. This delayed failure during shear deformation provides a $\approx$35\% improvement in the ideal strain tolerated by the structure, which contributes to the ductile nature of Mo$_2$BC.

\begin{figure*}[]
\includegraphics[width=6in]{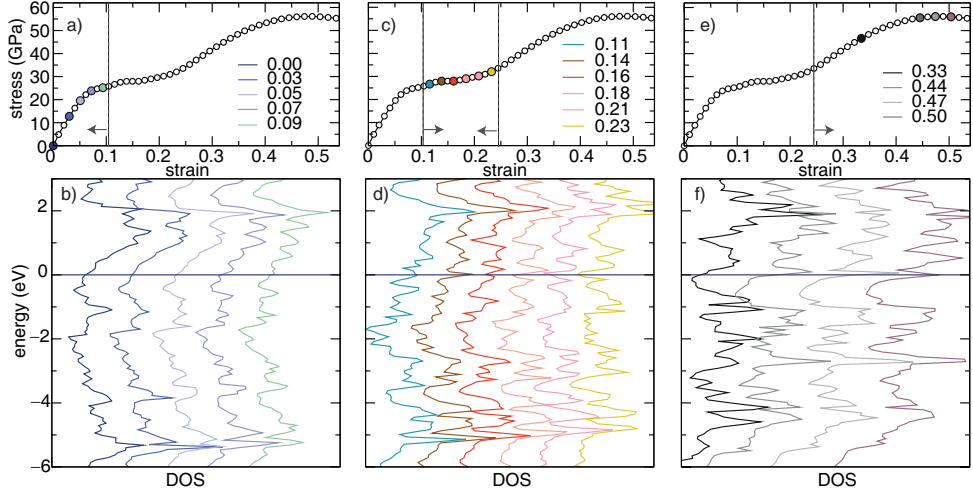} 
\centering
\caption{ The stress-strain curve for [001] in Mo$_2$BC is segmented into three interdependent stages. a) Stage 1 of the stress-strain curve is plotted with the highlighted strain steps corresponding  b) the density of state (DOS) of Stage 1. c) Shows the Stage 2 with the color-coded points corresponding to d) the DOS of Stage 2. e) Stage 3 shows the final stage of the stress-strain curve and the colors correspond to f) different strain steps in the DOS of Stage 3.} 
\label{fig:2}
\end{figure*}

The mechanistic origin of this unique strain-stiffening behavior was resolved by examining the electronic structure of Mo$_2$BC under mechanical perturbation. First, the [001] tensile direction, which could be a key component of the mechanical properties in Mo$_2$BC, was investigated. Plotting the [001] stress-strain curve in Fig.\,\ref{fig:2} shows it can be divided into three distinct stages (up to 0.53 strain). The curve's first stage occurs between strains 0-0.1, the second stage is between strains 0.1-0.25, and the third stage falls from strains 0.25-0.5. The corresponding density of states (DOS) at multiple strain steps for each stage have also been plotted and color-coded in Fig.\,\ref{fig:2} to illustrate the electronic structure perturbations as the structure is strained. Fig.\,\ref{fig:2}a shows that in the first stage, the stress increases almost linearly as the compound is strained, which follows conventional stress-strain behavior of crystalline metals and ceramics with a parabolic work-hardening.\cite{Meyers2009} The corresponding DOS curve shows the metallic behavior of Mo$_2$BC with the Fermi level (E$_f$) residing on a shoulder, dominated by 4$d$ orbitals of Mo as shown by Fig.\,S3. in supporting information.

As the compound is strained, the DOS undergoes minor changes with E$_f$ shifting to lower energy owing to differences in the bond lengths followed by unit cell volume expansion. As the compound is further strained in the final steps of the first stage, a small local maximum appears in the DOS at E$_f$ (strain 0.14 in Fig.\,\ref{fig:2}b) indicating a potential electronic instability. At this point, a dramatic decrease in the stress-strain curve would be expected, and the material should fail as shown by a drop in the stress. Instead of failure, the curve merely flattens as the compound enters the second strain stage (Fig.\,\ref{fig:2}c). This process coincides with the formation of a pseudogap at E$_f$ for 0.16 strain, shown in Fig.\,\ref{fig:2}d. The structure overcomes the energetic barrier imposed from the strain through the intrinsic formation of an electronically stabilizing pseudogap. This prevents the anticipated failure of Mo$_2$BC at this point and allows the structure to accommodate additional strain. The pseudogap in the DOS remains through the end of the second stage of the stress-strain curve. Proceeding to the third stage, illustrated in Fig.\,\ref{fig:2}e, shows the stress starts to increase again until reaching the ideal stress and strain at 56 GPa and 0.47, respectively, (Fig.\,\ref{fig:2}c). Correspondingly, E$_f$ moves to the shoulder of the pseudogap in DOS (Fig.\,\ref{fig:2}f) prior to falling on a peak at strain 0.50 at which point the compound is mechanically and electronically unstable with no possibility for a second recovery. The formation of a pseudogap in the DOS during the second stage provides mechanistic support for the observed intermediate strain-stiffening. Nevertheless, this phenomenon is just one component of the simultaneous high hardness and ductility in Mo$_2$BC. 

\begin{figure}[h] 
\includegraphics[width=3in]{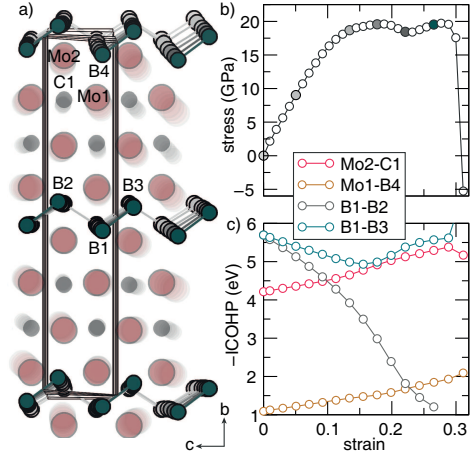}
\centering
\caption{a) Stacked snapshots of the crystal structure of Mo$_2$BC during (111)[$\bar{1}\bar{1}$2] deformation at shear strain steps of 0, 0.05, 0.13, 0.18, 0.22, and 0.26.  b) Stress-strain behavior of Mo$_2$BC under (111)[$\bar{1}\bar{1}$2] shear deformation and c) $-$ICOHP values as a function of strain during (111)[$\bar{1}\bar{1}$2] shear for various interactions.} 
\label{fig:3}
\end{figure}

Further unraveling the mechanism of failure in Mo$_2$BC requires also understanding the anomalous two-step behavior of the (111)[$\bar{1}\bar{1}$2] shear stress-strain curve. Analyzing the DOS curves for the (111)[$\bar{1}\bar{1}$2] shear system (Fig.\,S2 in supporting information) does not indicate the similar opening of a pseudogap. Instead, E$_f$ remaining on the shoulder of a peak in the DOS at all strains examined suggesting a different stabilizing mechanism. Therefore, it is necessary to take a closer look at the changes in the crystal structure and chemical bonding during the deformation. Fig.\,\ref{fig:3}a depicts superimposed snapshots of the Mo$_2$BC crystal structure under (111)[$\bar{1}\bar{1}$2] shear deformation with increasing strain. These snapshots correspond to the strain steps highlighted in Fig.\,\ref{fig:3}b and demonstrate the evolution of the unit cell, and atomic coordinates along the deformation path. Under the shear strain, the structure's symmetry breaks allowing the B-B interactions to split into two nonequivalent B1-B2 and B1-B3 interactions. 
Initially, under no strain, the boron-boron bond lengths are all equal and separated by 1.83\,\AA. As the crystal structure is strained the B1-B2 interatomic distance increases from 1.87\,\AA\ at 0.05 strain to 2.49\,\AA\ at 0.26 strain. This distance is sufficient to consider the bond effectively broken. However, at the same time, the B1-B3 interaction actually undergoes a decrease in bond length shortening from 1.85\,\AA\ at 0.05 strain to 1.83\,\AA\ at 0.26 strain. This asymmetric progression of the bond lengths is representative of dimerization of the boron-boron interactions, which is visualized in Fig.\,\ref{fig:3}a with the increase in B1-B2 interatomic distance indicated by the absence of a bond at strains $>$0.13 whereas the B1-B3 bond remains for all strain steps. 

A quantitative analysis of changes among the interatomic interactions can be made by decomposing the DOS into the bonding and antibonding components based on a crystal orbital Hamilton population ($-$COHP) analysis (Fig.\,S2 in supporting information).\cite{Dronskowski1993, Maintz2016} Evaluating the evolution of the interatomic interactions along the (111)[$\bar{1}\bar{1}$2] shear deformation path, the integrated $-$COHP ($-$ICOHP) values, which tend to scale with the strength of the interactions, were calculated and are plotted as a function of strain in Fig.\,\ref{fig:3}c. From these data it is clear the loss of symmetry allows the B-B interactions to independently change leading to substantial changes in the chemical bonding with the most notable being a $\approx$79\,\% decrease of the $-$ICOHP values for the B1-B2 interaction at maximum strain (Fig.\,\ref{fig:3}c). Exploring the fluctuations in the other $-$ICOHP values during the shear deformation makes it clear the B1-B2 contact weakens the most. Perturbing the structure further, causes the B1-B2 contact to continue a nearly linear decrease to only $\approx$1 eV/bond. Such a dramatic loss of covalent bonding character usually is sufficient to cause a structure to fail. However, analyzing the B1-B3 $-$ICOHP value shows this drops by only 0.76\,eV/bond, which is only half of the 2.19\,eV/bond decrease calculated for the B1-B2 interaction up to a strain of 0.18. Moreover, the $-$ICOHP value for B1-B3 contact under additional strain unexpectedly begins to increase with greater strain reaching a maximum of 5.5\,eV/bond at strain 0.26. This is nearly equivalent to the unstrained crystal structure (5.69 eV/bond). The transition of the B1-B3 $-$ICOHP value from decreasing, indicating a weakening of the interaction, to increasing, indicating a strengthening of the interaction, is directly correlated with the structure's mechanical properties.

From this analysis, it is clear that the first drop in the stress-strain curve occurs as the boron zigzag chain begins to dimerize. The B1-B3 dimer, however, beings to strengthen as implied by a shortening of the bond length as well as an increase in $-$ICOHP value. Moreover, Fig.\,\ref{fig:3}c also highlights an increase in the $-$ICOHP values for the Mo2-C1 and Mo1-B4 contacts by $\approx$25\,\% and $\approx$70\,\%, respectively, as the structure is strained. It is also worth noting that the $-$ICOHP values are more significant for the Mo-C interactions compared to the Mo-B interactions, in agreement to the more substantial hybridization of Mo-C 4$d$-2$p$ orbitals versus Mo-B 4$d$-2$p$ orbital, as determined from the partial DOS (Fig\,S1 in supporting information). 
This culmination of changes in the electronic structure leads to the formation of an electronically metastable state for Mo$_2$BC with chemical bonds that are actually enhanced by straining the crystal structure.

In summary, the complex crystal chemistry and stress-strain behavior in Mo$_2$BC supports the compound's high hardness while also bearing surprising ductility. This unusual mechanical response stems from an observed strain stiffening, which is explained through the formation of an electronic pseudogap within the DOS and dimerization of the boron-boron zigzag interactions. The results show that the strength in Mo$_2$BC does not merely rely on the sum of bond strength and is more complicated. The tensile strain-stiffening along [001] enhances its ultimate strength and strain while the sequential bond breakage and delayed failure along (111)[$\bar{1}\bar{1}$2] contribute to the unexpected ductility. The insight gained from this study not only provides a mechanistic answer for the unique mechanical response of Mo$_2$BC, but it also provides a better understanding of the complex and relatively unexplored intrinsic mechanical behavior of transition-metal borocarbides and more specifically crystalline solids with intriguing and uncommon mechanical properties.

\section{Acknowledgments}

The authors gratefully acknowledge the generous financial support provided by the National Science Foundation through No. NSF-CMMI 15-62142 and the donors of the American Chemical Society Petroleum Research Fund (55625-DNI10) for supporting this research.
Support for this work was provided by resources of the uHPC cluster managed by the University of Houston and acquired through NSF Award Number 15-31814. This research also used the Maxwell and Opuntia Clusters operated by the University of Houston and the Center for Advanced Computing and Data Systems.

\bibliography{Mo2BC_strength}

\end{document}


\sloppy

\title{Supporting information for: "An unconventional mechanism for simultaneous high hardness and ductility in Mo$_{2}$BC"}

\author{Aria Mansouri Tehrani}
\affiliation{Department of Chemistry, University of Houston, Houston, TX, USA}
\author{Amber Lim}
\affiliation{Department of Chemistry, University of Houston, Houston, TX, USA}
\author{Jakoah Brgoch}
\email{jbrgoch@uh.edu}
\affiliation{Department of Chemistry, University of Houston, Houston, TX, USA}
\affiliation{Texas Center for Superconductivity, Houston, TX, USA}


\keywords{}%
\maketitle 


\begin{figure}[h] 
\includegraphics[width=3in]{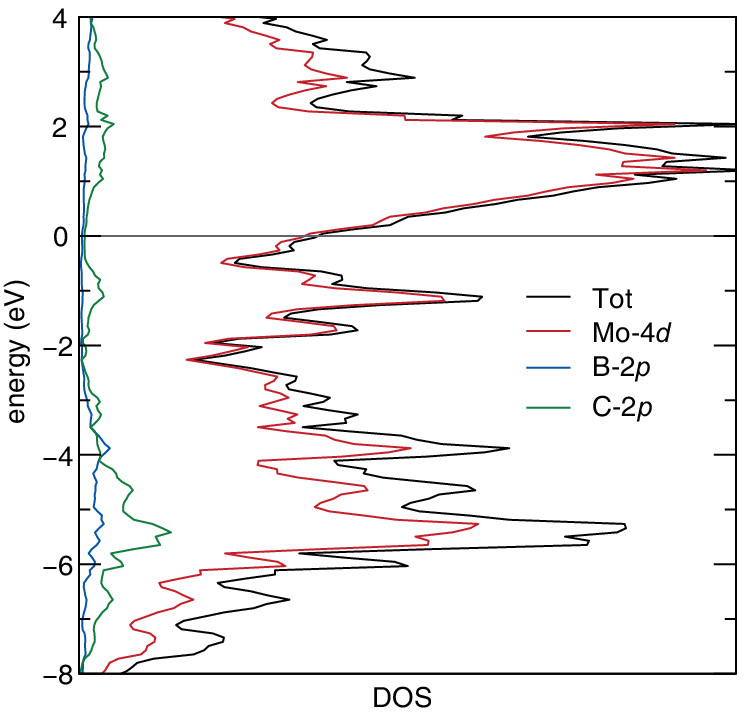}
\centering
\caption{Partial DOS of unstrained Mo$_2$BC.} 
\label{fig:3}
\end{figure}

\begin{figure}[h]
\includegraphics[width=3in]{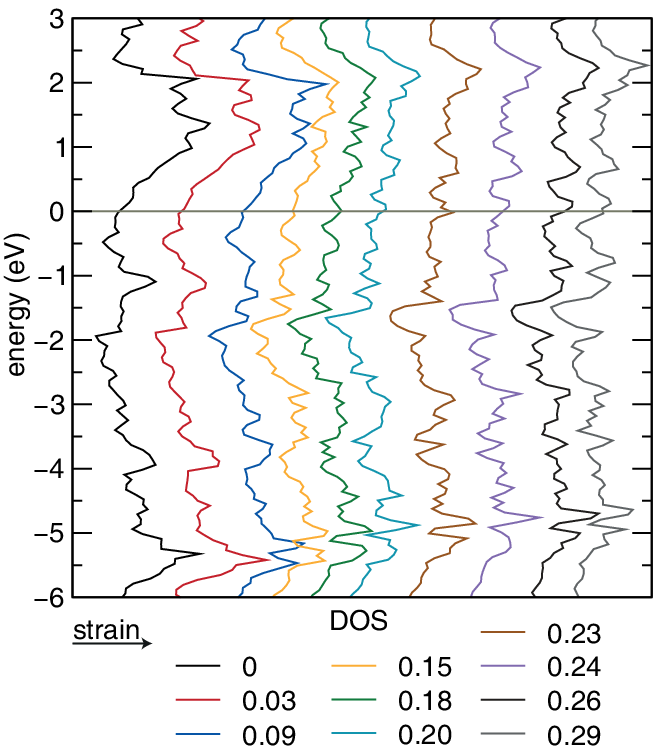}
\centering
\caption{DOS evolution as a function of strain for the shear deformation along (111)[$\bar{1}\bar{1}$2].} 
\label{fig:1}
\end{figure}

\begin{figure}[h]
\includegraphics[width=3in]{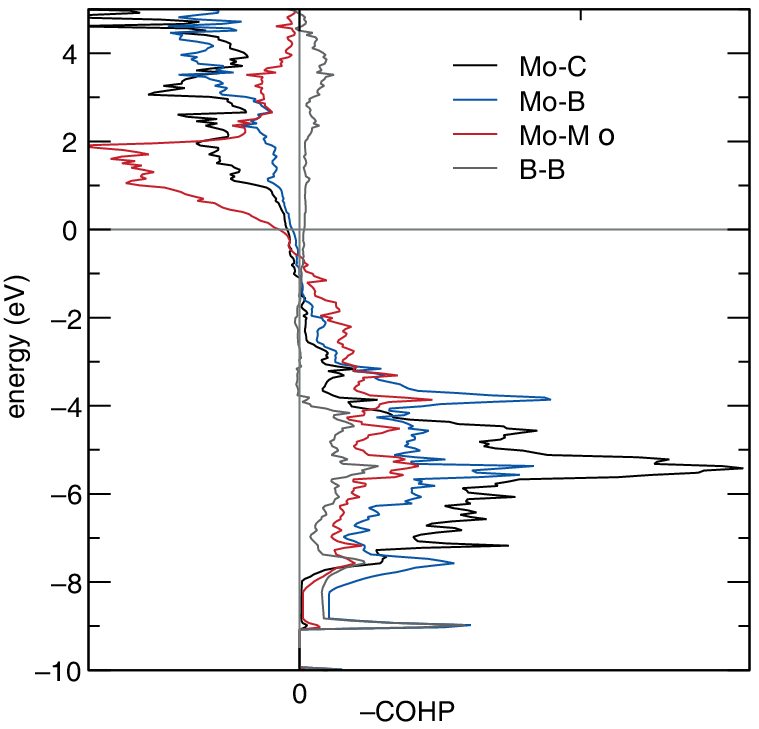} 
\centering
\caption{$-$COHP curves for various interatomic interaction in unstrained Mo$_2$BC.} 
\label{fig:2}
\end{figure}

\begin{table}[!h]
\caption{Ideal stress and strain values for various shear systems derived from stress-strain curves.}
\centering
\label{my-label}
\begin{tabular}{ccc}
\hline
applied shear    & ideal stress (GPa) & ideal strain (GPa) \\
\hline
(001){[}-100{]}  & 37.0               & 0.21               \\
(001){[}-110{]}  & 27.7               & 0.19               \\
(001){[}010{]}   & 23.5               & 0.19               \\
(001){[}100{]}   & 38.5               & 0.25               \\
(001){[}110{]}   & 27.7               & 0.19               \\
(010){[}-100{]}  & 22.9               & 0.17               \\
(010){[}001{]}   & 23.5               & 0.19               \\
(010){[}100{]}   & 22.9               & 0.17               \\
(010){[}101{]}   & 24.0               & 0.28               \\
(011){[}01-1{]}  & 26.0               & 0.17               \\
(011){[}1-11{]}  & 23.9               & 0.17               \\
(011){[}100{]}   & 26.5               & 0.17               \\
(011){[}11-1{]}  & 28.3               & 0.17               \\
(100){[}0-11{]}  & 26.5               & 0.17               \\
(100){[}001{]}   & 38.5               & 0.25               \\
(100){[}010{]}   & 22.9               & 0.17               \\
(100){[}011{]}   & 26.5               & 0.17               \\
(110){[}001{]}   & 27.4               & 0.17               \\
(110){[}1-10{]}  & 23.7               & 0.21               \\
(110){[}1-11{]}  & 30.7               & 0.42               \\
(110){[}-110{]}  & 50.2               & 0.19               \\
(111){[}-1-12{]} & 20.3               & 0.32               \\
(111){[}-101{]}  & 23.1               & 0.37               \\
(111){[}-110{]}  & 25.8               & 0.19               \\
(111){[}0-11{]}  & 23.8               & 0.17               \\
(111){[}01-1{]}  & 28.3               & 0.17               \\
(111){[}11-2{]}  & 27.8               & 0.28                \\
\hline
\end{tabular}
\end{table}